\documentclass[useAMS,usenatbib]{mn2e}
\usepackage{graphicx, color}
\usepackage{aas_macros}

\usepackage{braket}
\usepackage{bm}
\usepackage{threeparttable}
\usepackage{amssymb, amsmath}

\newcommand{\Msun}{\rm M_{\odot}}

\newcommand{\Lya}{Ly{$\rm \alpha$}~}
\newcommand{\Ha}{H{$\rm \alpha$}~}
\newcommand{\Hb}{H{$\rm \beta$}~}
\newcommand{\OIII}{O\,{\sc iii}}
\newcommand{\HII}{H\,{\sc ii}~}
\newcommand{\CII}{C\,{\sc ii}}

\title[21-cm vs \OIII~emitters]
{Cross-correlation between the 21-cm signal and [\OIII] emitters during early cosmic reionization}

\author[K. Moriwaki et al.]
{Kana Moriwaki$^{1}$\thanks{E-mail: kana.moriwaki@utap.phys.s.u-tokyo.ac.jp},
Naoki, Yoshida$^{1, 2, 3}$,
Marius B. Eide$^{4}$,
Benedetta Ciardi$^{4}$
\\
$^{1}$Department of Physics, The University of Tokyo, 
7-3-1 Hongo, Bunkyo, Tokyo 113-0033, Japan \\
$^{2}$Kavli Institute for the Physics and Mathematics of the Universe (WPI), 
UT Institutes for Advanced Study, \\
The University of Tokyo, 5-1-5 Kashiwanoha, Kashiwa, Chiba 277-8583, Japan \\
$^{3}$Research Center for the Early Universe, School of Science, The University of Tokyo, 
7-3-1 Hongo, Bunkyo, Tokyo 113-0033, Japan \\
$^{4}$Max-Planck-Institut f\"{u}r Astrophysik, Karl-Schwarzschild-Stra\ss e 1, D-85741 Garching, Germany
}

\begin{document}

\date{}

\pagerange{\pageref{firstpage}--\pageref{lastpage}}
\pubyear{0000}

\maketitle

\label{firstpage}

\begin{abstract}
We study statistics of the 21-cm signal from the epoch of reionization.
We propose to use [\OIII] line emitting galaxies to cross-correlate
with the 21-cm signal from $z = 7 - 10$. 
To this aim, we employ simulations of reionization obtained post-processing the high-resolution cosmological hydrodynamics simulation Massive Black-II with the three-dimensional (3D) radiative transfer code {\sc crash} to follow the propagation of ionizing photons from a variety of sources.
We show that, during the early phases of reionization, the 21-cm signal is positively correlated with the spatial distribution of the [\OIII] emitters on large scales $(k < 1h~\rm cMpc^{-1})$.
This positive correlation is generated by the temperature - galaxy correlation and it is a few times larger than when we assume that the heating is saturated.
As the reionized regions expand, the correlation changes its sign to negative from $z = 10$ to 8. 
The signals at this epoch can be detected by combining the Square Kilometre Array (SKA) and a wide-field [\OIII] emitter survey.
We also calculate the cross-power spectrum with a 3D [\OIII] intensity field, aiming at exploiting future intensity mapping observations.
We conclude that high-redshift [\OIII] line emitters can be used to probe the reionization process when the inter-galactic medium is largely neutral. 
\end{abstract}

\begin{keywords}  
 galaxies: high-redshift --
 intergalactic medium --
 dark ages, reionization, first stars
\end{keywords}

\section{INTRODUCTION}

A number of observational programmes are planned to probe the evolution of the intergalactic medium (IGM) during cosmic reionization. So far, observations of absorption lines in quasar spectra (e.g. \citealt{McGreer15}) 
and the polarisation of the cosmic microwave background \citep{Planck15} suggest
that the IGM was almost fully ionized  by $z \sim 6$, whereas the recent global hydrogen 21-cm signal measurement by EDGES, if confirmed, indicates that reionization began as early as $z\sim 15$ \citep{Bowman18}.
Although the inferred epoch of reionization (EoR) is roughly consistent with theoretical models based on the
standard cosmology,
there still remain outstanding issues such as the nature of its sources and the detailed ionization history.

The hydrogen 21-cm line provides a direct method to probe the distribution of neutral hydrogen in the Universe. 
Ongoing and planned radio observations include 
Murchison Wide Field Array (MWA)\footnote{http://www.mwatelescope.org},
Low Frequency Array (LOFAR)\footnote{http://www.lofar.org}, 
Precision Array for Probing Epoch of Reionization (PAPER)\footnote{http://eor.berkeley.edu},
Hydrogen Epoch of Reionization Array (HERA)\footnote{https://reionization.org},
and Square Kilometer Array (SKA)\footnote{https://astronomers.skatelescope.org}.
It is well known that a variety of the so-called foregrounds hamper robust detection of the 21-cm signal from the EoR. Most notably the Galactic synchrotron emission and extragalactic radio sources are far brighter than the expected 21-cm signal.
A promising way of mitigating the systematic noise or contamination from foregrounds, in addition to confirming the origin of the signal, is to utilise the cross-correlation of the 21-cm line with some other known tracers at the same redshift. 
The spatial cross-correlation between the 21-cm signal and the distribution of galaxies at the EoR has been suggested as one of such probes, and correlation with \Lya emitters (LAEs) has been studied extensively (e.g. \citealt{Lidz09, Wiersma13, Vrbanec16, Hutter17, Kubota18}).
While LAEs can be an excellent tracer of the underlying large-scale structure at $z \sim 6 - 7$, it becomes progressively difficult to detect LAEs toward higher redshift, when the neutral fraction of the IGM is large.
\citet{Konno14} has shown that the \Lya luminosity becomes significantly weak compared to UV continuum at $z > 7$.
It is thus necessary to investigate the possibility of using other tracers, if they exist, to measure the cross-correlation during the early phase of reionization.

Recent observations by Atacama Large Millimeter/submillimeter Array (ALMA) demonstrated that the FIR [\OIII] line is an excellent target to detect and study
galaxies at $z > 7$ (e.g. \citealt{Inoue16, Carniani17, Laporte17, Hashimoto18a, Hashimoto19, Tamura19}).
Earlier in our study, we propose that [\OIII] line emitters can be used as a tracer of the underlying large-scale structure
at high redshift \citep{Moriwaki18}.
Line intensity mapping is another powerful method to study the large-scale distribution of galaxies.
Intensity mapping observations of the high-redshift [\CII], CO, \Ha  lines are already planned \citep{Kovetz17}.
The Spectrophotometer for the History of the Universe, Epoch of Reionization, and Ice Explorer (SPHEREx), which is aiming to map the \Ha intensity at $z < 6$, can also detect the high-redshift optical [\OIII] line in the future.
The possibility of using intensity mapping, rather than galaxy surveys,
in cross-correlation studies should then be considered, and indeed has been proposed by a number of authors (e.g. \citealt{Visbal10,Silva15,Dumitru19}).

In this paper, we examine the cross-correlation between the 21-cm line and [\OIII] emitters.
The paper is structured as follows. In Section~2 we describe the methodology used for this study, in Section~3 we present the results, and in Section~4 we discuss future prospects for the detectability of the correlation and give our conclusions.

\begin{figure*}
\begin{center}
\includegraphics[width=13.5cm]{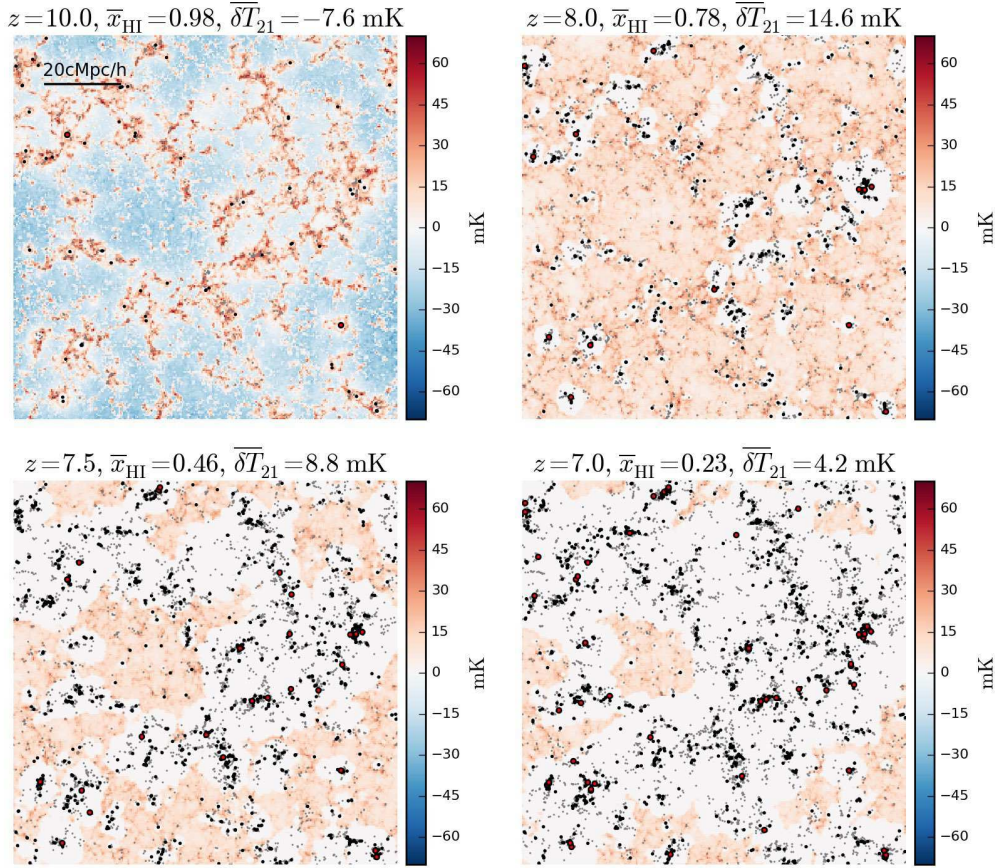}
\caption{
Slice maps of the 21-cm signals at $z = 10.0$,  8.0,  7.5, and 7.0.
The black and red points represent the galaxies with $L_{\rm O_{III}} > 10^{41}~\rm erg~s^{-1}$ and $10^{42}~\rm erg~s^{-1}$ respectively and the gray points are all the galaxies within the slice.
The width of the slices is $1.2h^{-1}~\rm cMpc$.
}
\label{fig:map_21_gal}
\end{center}
\end{figure*}

\section{METHODOLOGY}

In this section we present the simulation and the methodology employed to evaluate the cross-correlation between the 21-cm and [\OIII] lines.

\subsection{Simulations of reionization}

Here we provide a brief description of the simulations of reionization employed in this work, while we refer the reader to  \citet{Eide18} and Eide et al. in prep for more details.

The IGM and source properties are retrieved from Massive Black-II, a high-resolution cosmological hydrodynamics simulation of galaxy formation \citep{Khandai15}, which follows physical processes such as star formation, feedback, and black hole (BH) evolution \citep{Springel03, DiMatteo05, Springel05, Croft09, Degraf10, DiMatteo12}.
The simulated volume is a cube of $100h^{-1}$ cMpc on a side, where $2 \times 1792^3$ gas and dark matter particles with masses of $m_{\rm gas} = 2.2\times 10^6h^{-1}~\Msun$ and $m_{\rm DM} = 1.1\times 10^7h^{-1}~\Msun$ are distributed initially, so that the minimum halo mass identified in the simulation is $\sim 10^8~\Msun$.
The cosmological parameters are from the 7-year Wilkinson Microwave Anisotropy Probe (WMAP7), i.e. $\sigma_8 = 0.816$, $n_s = 0.968$, $\Omega_\Lambda = 0.725$, $\Omega_{\rm m} = 0.275$, $\Omega_{\rm b} = 0.046$ and $h=0.701$ \citep{Komatsu11}.

Radiation transfer (RT) is done as in \citet{Eide18}. Briefly,
a density and a source distribution map on $256^3$ cells are generated from each snapshot.
The Monte Carlo ray-tracing code {\sc crash} \citep{Ciardi01, Maselli03, Maselli09, Graziani13, Graziani18}
is then used to compute the ionization degree and the gas temperature of each cell.
Four types of ionizing sources are considered: stars, X-ray binaries, supernova-heated interstellar medium, and nuclear BHs.
A constant escape fraction of UV photons $f_{\rm esc} = 0.15$ is assumed for all the sources except BHs, for which the escape fraction is unity.  
Further details on the spectral shapes and the ionizing photon production rates are found in \citet{Eide18} and Eide et al. in prep.
The reionization simulation yields the Thomson scattering optical depth of the cosmic microwave background (CMB) $\tau_{\rm e} = 0.055$, which is consistent with the Planck result \citep{Planck18}.
We note here that stars contribute to generate and heat fully ionized H{\sc ii} regions, whereas the other, more energetic sources contribute to partial ionization and heating of the diffuse IGM. 

\subsection{21-cm signal}

The 21-cm brightness temperature relative to the CMB temperature is given by (e.g. \citealt{Field59, Furlanetto06})
\begin{eqnarray}
	\delta T_{21} = T_0 \, \overline{x}_{\rm HI}\,(1+\delta_x)(1+\delta_\rho) \Big( \frac{T_{\rm s} - T_{\rm CMB}}{T_{\rm s}}\Big),  
	\label{eq:21-cm}
\end{eqnarray}
where $\overline{x}_{\rm HI}$ is the volume averaged neutral fraction, $T_{\rm s}$ is the spin temperature, 
$\delta_x = x_{\rm HI} / \overline{x}_{\rm HI} - 1$ and $\delta_\rho = \rho / \overline{\rho} - 1$ are 
the fluctuations of the neutral fraction and the gas density respectively, and
\begin{eqnarray}
	T_0 = 28.5\,\Big(\frac{\Omega_{\rm b}h^2}{0.022}\Big)\Big(\frac{1+z}{10}\frac{0.15}{\Omega_{\rm m}h^2}\Big)^{1/2} {\rm ~mK}
\end{eqnarray}
is the normalization factor that scales with cosmological parameters.
The spin temperature is set to be the gas temperature, which is a valid assumption in the presence of strong \Lya coupling. We also make the common assumption that $(1 + \frac{1}{H(z)} \frac{{\rm d}v_\parallel}{{\rm d}r_\parallel})$ $\sim 1$, where $H(z)$ is the Hubble parameter and ${\rm d}v_{\parallel}/{{\rm d}r_\parallel}$ is the velocity gradient along the line of sight.
To examine the contribution from the temperature fluctuations to the 21-cm signal, we rewrite equation (\ref{eq:21-cm}) as
\begin{eqnarray}
    \delta T_{21} = T_0 \, \overline{x}_{\rm HI} \, \overline{\eta}\,(1+\delta_x)(1+\delta_\rho)(1+\delta_\eta),
    \label{eq:21-cm_eta}
\end{eqnarray}
by introducing $\eta \equiv 1-T_{\rm CMB}/T_{\rm s}$, where $\overline{\eta}$ is the volume averaged value of $\eta$ and $\delta_\eta = \eta / \overline{\eta} - 1$. 

Fig. \ref{fig:map_21_gal} shows the slice maps of the 21-cm signals
at $(z, \overline{x}_{\rm HI}) = (10.0, 0.98)$, $(8.0, 0.78)$, $(7.5, 0.46)$, and $(7.0, 0.23)$.
The red (blue) regions correspond to areas where the emission (absorption) signals are seen.
At $z = 10$, the gas temperature in low density regions distant from bright galaxies is smaller than the CMB temperature, and thus absorption features are clearly seen in such regions. 
For a more extensive discussion of the 21-cm signal obtained from the simulations we refer the reader to Ma et al. in prep.

\subsection{[\OIII] line emission}

\begin{table}
\caption{
Redshift, volume averaged neutral fraction, volume averaged luminosity density, 
and number of galaxies with [\OIII] luminosity larger than $10^{41}$ and $10^{42} \rm~ erg~s^{-1}$ within $(100h^{-1}~\rm cMpc)^3$ for each snapshot.
}
\centering
\begin{threeparttable}
	\begin{tabular}{ccccc} \hline
	redshift &
	$\overline{x}_{\rm HI}$ &
	$\overline{l}_{\rm OIII}$\tnote{a}& 
	$N_{\rm gal,41}$ & 
	$N_{\rm gal,42}$ \\
	\hline 
	10.0 & 0.98 & 0.197 & 4851  & 84    \\
	8.0  & 0.78 & 1.13  & 30399 & 856   \\
	7.5  & 0.46 & 2.66  & 72619 & 2522  \\
	7.0  & 0.23 & 3.53  & 92593 & 3634  \\ \hline
	\end{tabular}
\label{table:snapshots}
\begin{tablenotes}
\item [a] in units of $10^{40}h^{3}~\rm erg~s^{-1}~cMpc^{-3}$.
\end{tablenotes}
\end{threeparttable}
\end{table}

FIR/optical [\OIII] lines are excellent targets to study the high-redshift large-scale structure \citep{Moriwaki18}.
Emission lines of highly ionized heavy elements including [\OIII] lines originate from \HII regions around young and massive stars, and thus, they directly trace the ionizing sources.
In addition, they are easier to model compared to those emitted from both neutral and ionized regions, such as [\CII] line. 
Among them, the [\OIII] 5007\AA~line is one of the easiest to model because the electron density in \HII regions, which cannot be properly resolved in simulations as large as those adopted here, does not affect its luminosity. 
In this paper, we thus investigate the [\OIII] 5007\AA~line.
If we considered the FIR [\OIII] line instead, the detectability would change depending on the emission strength but the overall results would still hold as long as the electron densities do not vary substantially from galaxy to galaxy.

To compute the line luminosity, $L_{\rm O_{III}}$, we use a library generated with the photoionization code {\sc cloudy} \citep{Ferland17} as in \citet{Moriwaki18}, which contains the line luminosity relative to the \Hb luminosity with the case-B approximation, $L_{\rm H\beta}^{\rm case B}$.
The [\OIII] 5007\AA~line luminosity is calculated as
\begin{eqnarray}
	L_{\rm O_{III}} = (1-f_{\rm esc}) \, C_{\rm O_{III}}(Z, U, n) \, L_{\rm H\beta}^{\rm case B},
	\label{eq:oiii_luminosity}
\end{eqnarray}
where $C_{\rm O_{III}}$ is the line luminosity ratio calculated with {\sc cloudy}, $Z$ is the mean gas metallicity of a galaxy, $U$ is the ionization parameter, and $n$ is the electron density.
We calculate the ionization parameter as (e.g. \citealt{Panuzzo03})
\begin{eqnarray}
	U = \frac{3\alpha_{\rm B}^{2/3}}{4c}\Big(\frac{3\dot{N}_{\rm ion} n}{4\pi}\Big)^{1/3},
\end{eqnarray}
where $\alpha_{\rm B}$ is the case-B hydrogen recombination coefficient, and $\dot{N}_{\rm ion}$ is the number of ionizing photons emitted from sources other than BHs multiplied by $(1-f_{\rm esc})$.
As already mentioned, the [\OIII] 5007\AA~line luminosity is typically independent from the electron density because its critical density is as high as $10^6~\rm cm^{-3}$, while the typical electron density in \HII regions is $10 - 1000~\rm cm^{-3}$ (e.g. \citealt{Sanders16}).
Because the simulation does not resolve individual \HII regions within a galaxy, we adopt $n =100~\rm cm^{-3}$ for all galaxies in this study, but we verified that the results are not affected by this specific choice.

For each output, we generate a number density map of [\OIII] emitters, $n_{\rm gal}(\bm r)$, and a [\OIII] intensity map, $I_{\rm O_{III}}(\bm r)$, with $256^3$ cells to cross-correlate with the 21-cm signal.
The [\OIII] intensity at the position $\bm r$ is given by
\begin{eqnarray}
    I_{\rm O_{III}}(\bm r) = \frac{L_{\rm O_{III}}(\bm r)}{4\pi D_{\rm L}^2} \frac{\chi^2\frac{{\rm d}\chi}{{\rm d}\lambda}}{V_{\rm cell}} 
    = \frac{\chi^2\frac{{\rm d}\chi}{{\rm d}\lambda}}{4\pi D_{\rm L}^2} l_{\rm O_{III}}(\bm r), 
\end{eqnarray}
where $L_{\rm O_{III}}(\bm r)$ is the total [\OIII] luminosity within a cell at $\bm r$,
$D_{\rm L}$ is the proper luminosity distance, 
$\chi$ is the comoving distance, 
$\lambda$ is the observed wavelength, 
$V_{\rm cell}$ is the comoving volume of a cell,
and $l_{\rm O_{III}} (\bm r)$ is the comoving luminosity density at $\bm r$.
Table \ref{table:snapshots} summarizes the global properties of each snapshot.
The [\OIII] luminosity density and the number of bright galaxies rapidly increase from $z = 10$ to $z = 8$, 
but the IGM neutral fraction remains larger than 0.7 at $z = 8$.
The distribution of galaxies is also shown in Fig. \ref{fig:map_21_gal}.
Bright [\OIII] emitters are embedded in large \HII regions 
as they are tracers of high density regions, where ionizing sources are expected to reside.

\subsection{Cross-power spectrum}

We calculate the cross-power spectrum between the 21-cm signal, $\delta T_{21}$, and the fluctuation of the galaxy number density, $\delta_n (\bm r) = 1 - n_{\rm gal}(\bm r)/\overline{n}_{\rm gal}$,
where $\overline{n}_{\rm gal}$ is the mean number density of the galaxies. 
In intensity mapping, the volume-averaged 
intensity over the observation volume, $\overline{I}_{\rm O_{III}}$, cannot be directly observed because of a variety of foreground sources and contamination.
Thus we cross-correlate the 21-cm signal with $I_{\rm O_{III}}(\bm r)$ instead of $\delta I_{\rm O_{III}}(\bm r) = 1 - I_{\rm O_{III}}(\bm r)/ \overline{I}_{\rm O_{III}}$.
We calculate the normalized three-dimensional (3D) cross-power spectra between $\delta T_{21}(\bm r)$ and $\delta_n (\bm r)$ ($\Delta^2_{21, n}$) or $I_{\rm O_{III}}(\bm r)$ ($\Delta^2_{21, I}$) as
\begin{eqnarray}
	\Delta^2_{21,i}(k) = \frac{k^3}{2\pi^2} P_{21, i}(k)~~~(i = n, I).
\end{eqnarray}
As a consequence of the three fluctuation terms ($\delta_x$, $\delta_\rho$ and $\delta_\eta$) in equation \eqref{eq:21-cm_eta}, the cross-power spectrum is composed of seven terms:
\begin{eqnarray}
    \Delta^2_{21,i} &=& T_0 \, \overline{x}_{\rm HI} \, \overline{\eta} \,
    ( \Delta^2_{x,i} + \Delta^2_{\rho,i} + \Delta^2_{\eta, i} \nonumber \\
    &&+ \Delta^2_{x\rho, i} + \Delta^2_{x \eta, i} + \Delta^2_{\rho \eta, i} + \Delta^2_{x\rho \eta, i} ).
\end{eqnarray}
If the IGM gas is much hotter than the CMB temperature (i.e., $\eta \sim 1$), the cross-power spectrum is reduced to
\begin{eqnarray}
        \Delta^2_{21,i} = T_0 \, \overline{x}_{\rm HI} \,
        (\Delta^2_{x,i} + \Delta^2_{\rho, i} + \Delta^2_{x\rho, i}),
\end{eqnarray}
as often assumed in previous studies.
In the following section, we will show the cross-power spectrum calculated under this "saturated heating assumption" to examine the effect of temperature fluctuations.

We also calculate the cross-correlation coefficient 
\begin{eqnarray}
	r_{21,i}(k) = \frac{P_{21, i}(k)}{ \sqrt{ P_{21}(k) P_i (k) }},
\end{eqnarray}
where $P_{21}(k)$ and $P_g (k)$ are the respective autopower spectra.
We subtract the shot noise after calculating the autopower spectrum $P_i (k)$.

\section{Results}

In this section we will discuss our results in terms of cross-correlation between the 21-cm signal and the [\OIII] emitters as well as [\OIII] intensity.

\subsection{21 cm-[\OIII] emitter cross-power spectrum}

\begin{figure}
\begin{center}
\includegraphics[width=7cm]{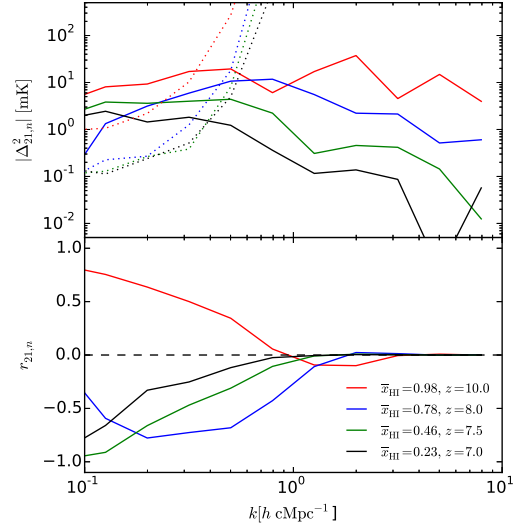}
\caption{
3D cross-power spectra between the 21-cm signal and the number density of [\OIII] emitters with $L_{\rm O_{III}} > 10^{42}~\rm erg~s^{-1}$ (top) and the corresponding cross-correlation coefficients (bottom) at $z = 10.0$ (red), 8.0 (blue), 7.5 (green) and 7.0 (black).
The dotted lines show the errors given in equation \eqref{eq:var} assuming SKA thermal noise and a hypothetical survey of [\OIII] emitters with a survey volume of $10^6 h^{-3}~\rm cMpc^3$ and a redshift error of 0.01. 
The logarithmic bin width $\Delta \log k$ is set to be 0.2.
}
\label{fig:cps.nden.42}
\end{center}
\end{figure}

We first calculate the 3D cross-power spectrum between the 21-cm signal and the fluctuation of the number density of galaxies.
We assume that all the galaxies with [\OIII] luminosity greater than $L_{\rm min} = 10^{42}~\rm erg~s^{-1}$
are detected.
\footnote{
For reference, NIRCam on James Webb Space Telescope can detect [\OIII] emitters with $L_{\rm O_{III}} > 2 - 3 \times 10^{42}~\rm erg~s^{-1}$ at $z = 7 - 9$ with S/N $> 5$ by integrating $10^4~\rm s$ per one field-of-view \citep{Moriwaki18}.}
Such galaxies are typically hosted by halos with masses greater than $M_{\rm halo} \sim 2\times 10^{10}$~M$_\odot$.
Fig. \ref{fig:cps.nden.42} shows the cross-power spectra at various redshifts (solid lines), together with the expected errors (dotted lines), which will be discussed later.

At $z = 10$, $\Delta^2_{\eta, n}$ dominates at $k < 1h~\rm cMpc^{-1}$, producing the positive correlation clearly seen in Fig. \ref{fig:cps.nden.42}. 
As our reionization model includes sources more energetic than stars, the gas temperature near the ionizing sources is saturated ($\eta \sim 1$) even at $z = 10$, and $\Delta^2_{\eta, n}$ disappears on scales smaller than the size of the saturated regions.
Instead, we observe an anti-correlation of $\Delta^2_{x\eta,n}$ on these small scales (i.e., $k\sim 2h~\rm cMpc^{-1}$). 
As the IGM gets heated with decreasing redshift, 
$\Delta^2_{\eta, n}$ gets smaller. 
Inversely, the anti-correlation of $\Delta^2_{x, n}$ becomes larger and shifts to larger scales as the ionized regions grow in size.
From $z = 10$ to 7.5, one can see a positive-to-negative transition of a large-scale cross-power spectrum.
At $z < 7.5$, we find that the IGM temperature is saturated almost everywhere and there is always anti-correlation on large scales. 

At every redshift, we observe another clear transition in the cross-correlation coefficient (the bottom panel of Fig. \ref{fig:cps.nden.42}) from anti-correlation to no-correlation on the scale corresponding to the typical size of fully ionized regions, i.e., $k = 3h~\rm cMpc^{-1}$ ($r = 2h^{-1}$ cMpc) at $z = 10$ and $k = 1h~\rm cMpc^{-1}$ ($r = 6h^{-1}$ cMpc) at $z = 7.5$. 
A positive correlation on these small scales has been found by other authors (see e.g. \citealt{Lidz09}) and it could be related to two different effects. 
If galaxies outside ionized regions are observed, then the correlation between the gas density and these galaxies generates positive correlation.
In our simulation, this is observed only when $L_{\rm min} < 10^{40}~\rm erg~s^{-1}$.
Alternatively, this could be associated to some residual neutral gas in high density regions within H{\sc ii} bubbles. In this case the neutral gas, and hence the associated 21-cm signal, is correlated with the galaxies residing within the same \HII region.
In our simulations, we find that the gas within \HII regions is fully ionized, as a result either of the contribution from a large number of small mass galaxies (not resolved e.g. in \citealt{Lidz09}) or of the lack of self-shielded Lyman limit systems as a consequence of the limited resolution of the simulation.
The cross-power spectra in Fig. \ref{fig:cps.nden.42} do not show clear negative to positive ``turnover''.

We next explore the detectability of the cross-power spectra.
The variance of cross-power spectrum for a particular mode $(k, \mu)$
is given by (e.g. \citealt{Lidz09})
\begin{eqnarray}
    \sigma^2_{21, n}(k,\mu) &=& \frac{1}{2} [P^2_{21,n}(k,\mu) + \sigma_{21}(k,\mu)\sigma_{n}(k,\mu)],
\end{eqnarray}
where $\mu$ is the cosine of the angle between $\bm k$ and the line of sight
and 
\begin{eqnarray}
    \sigma_{21}(k,\mu)  &=&  P_{21}(k,\mu) + P_{\rm N, 21}(k,\mu), \label{eq:err_21}\\
    \sigma_n (k,\mu) &=& P_n (k,\mu) + P_{{\rm N}, n}(k,\mu), \label{eq:err_gal}
\end{eqnarray}
are the variances of the autopower spectra.
The first terms, $P_{21}$ and $P_n$, represent sample variances and the second terms,
$P_{\rm N,21}$ and $P_{{\rm N},n}$, are the noise power spectra which
depend on the instruments (see equations (22) and (23) of \citealt{Kubota18}).
The variance of the spherically averaged cross-power spectrum is then 
calculated as
\begin{eqnarray}
    \frac{1}{\sigma^2_{{\rm 21}, n}(k)} = \sum_\mu \Delta \mu \frac{k^2\Delta k V_{\rm surv}}{4\pi^2} \frac{1}{\sigma^2_{21,n}(k, \mu)}, \label{eq:var}
\end{eqnarray}
where $V_{\rm surv}$ is the survey volume.
We assume a SKA survey \citep{Waterson16} with 1000 hour integration time, and a hypothetical redshift survey of [\OIII] emitters
capable of detecting galaxies with $L_{\rm O_{III}} > 10^{42}~\rm erg~s^{-1}$ with a redshift error $\sigma_z = 0.01$ and 
a survey volume $V_{\rm surv} = 10^6h^{-3}~\rm cMpc^3$,
which corresponds to an area of $1~\rm deg^2$ with a redshift interval of ${\rm d}z = 0.4$ at $z = 10$.
With this survey volume, a statistically significant number ($\sim 100$) of galaxies can be detected even at $z = 10$ (see table \ref{table:snapshots}).
For SKA observations, we adopt the system temperature, the number density of antenna tiles, and the effective area of each antenna tile from \citet{Kubota18}.
The dotted lines in Fig. \ref{fig:cps.nden.42} show the variance given in equation \eqref{eq:var}.
The large-scale correlation at $k \sim 0.1h~\rm cMpc^{-1}$ (and up to $k \sim 0.5h~\rm cMpc^{-1}$ depending on redshift) can be detected by combining SKA and our hypothetical galaxy survey.
We find that the sample variance ($P_{21}$ and $P_n $ terms in equations \eqref{eq:err_21} and \eqref{eq:err_gal}) dominate the errors at $k \sim 0.1h~\rm cMpc^{-1}$, whereas the observational errors ($P_{\rm N}$ terms) dominate on smaller scales.

\begin{figure}
\begin{center}
\includegraphics[width=7cm]{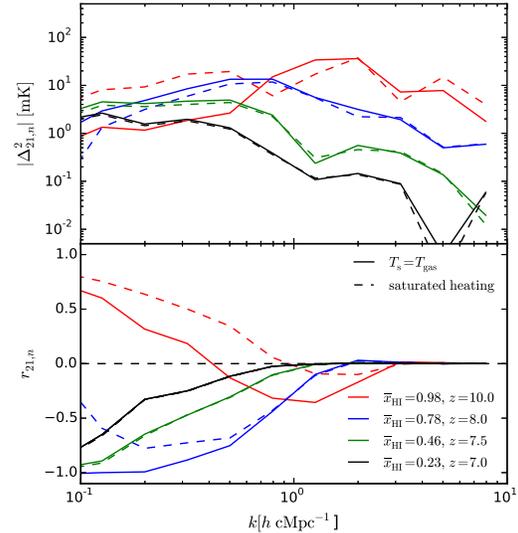}
\caption{
3D cross-power spectra (top) and the cross-correlation coefficients (bottom) between the 21-cm signal and the number density of [\OIII] emitters with $L_{\rm O_{\rm III}} > 10^{42}~\rm erg~s^{-1}$. We show the results in two cases; one with saturated heating assumption, $1 - T_{\rm CMB}/T_{\rm s} \sim 1$, shown by solid lines, and the other with $T_{\rm s} = T_{\rm gas}$, shown by dashed lines. The latter are identical to those in Fig. \ref{fig:cps.nden.42}.
}
\label{fig:saturated_heating}
\end{center}
\end{figure}

To examine the effect of temperature fluctuations, we calculate the cross-power spectra under the saturated heating assumption, i.e. $\eta \sim 1$.
The solid lines in Fig. \ref{fig:saturated_heating} show the cross-power spectra calculated with saturated heating assumption.
For reference, we also show the cross-power spectra calculated with the temperature fluctuations (i.e., $T_{\rm s} = T_{\rm gas}$, dashed lines), which are identical to those in Fig. \ref{fig:cps.nden.42}.
At $z = 10$, the positive correlation of $\Delta^2_{\rho, n}$ on large scales ($k < 0.4h~\rm cMpc^{-1}$) is present even without the fluctuations of $\eta$, but its amplitude is much smaller than $\Delta^2_{\eta, n}$.
At $z = 8$, we find stronger anti-correlation at $k < 1 h ~\rm cMpc^{-1}$ in the case with
the saturated heating assumption. This is owing to the absence of cancellation by $\Delta^2_{\eta, n}$
(see Equation [8]).
The $\eta$ fluctuations rapidly decay at $z < 10$, and the contribution to the 21cm cross-power gets smaller at lower redshifts.
Therefore, we expect a positive-to-negative transition in the cross-power spectrum at a similar redshift between $z = 8$ and 10 in both the cases, despite the large difference in the
cross-power amplitudes at $z = 10$.
We note that the heating rate of the IGM gas depends on the luminosity and the spectral shape of X-ray sources \citep{Fialkov14b, Fialkov17}.
One should keep in mind that the transition would occur at a later stage of the EoR if the IGM gas is heated more slowly, though the inefficient X-ray heating model is disfavored by recent observations (e.g. \citealt{Singh18, Monsalve19}).

\subsection{21 cm-[\OIII] intensity cross-power spectrum}

\begin{figure}
\begin{center}
\includegraphics[width=7cm]{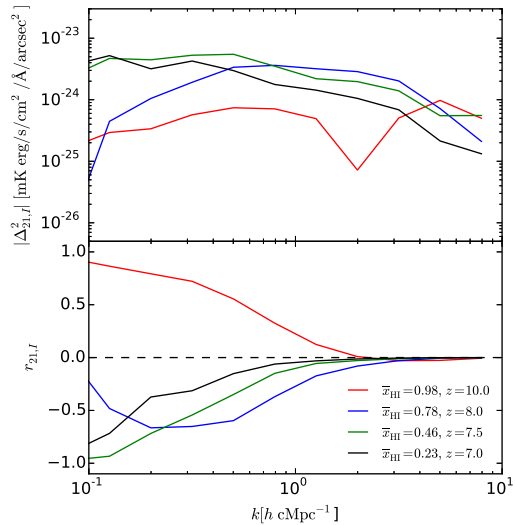}
\caption{3D cross-power spectra (top) and the cross-correlation coefficients (bottom) between 21-cm and [\OIII] intensity at $z=10.0$ (red), 8.0 (blue), 7.5 (green) and 7.0 (black). 
}
\label{fig:cps.lden}
\end{center}
\end{figure}

Next, we discuss the cross-correlation with [\OIII] intensity.
Fig. \ref{fig:cps.lden} shows the cross-power spectra between the 21-cm signal and the [\OIII] intensity at four redshifts. The positive-to-negative transition of cross-power spectrum occurs between $z = 10$ and 8, almost the epoch when $\Delta^2_{21,n}$ changes its sign.
The amplitude of the cross-power spectrum scales with the mean intensity. 
At $z = 10$, for instance, although there is a strong correlation between 21-cm and [\OIII] intensity on large scales, the cross-power amplitude is smaller than at lower redshifts because of the small mean intensity at $z = 10$ (see table \ref{table:snapshots}).
The transition in $r_{21,I}$ from anti-correlation to no-correlation is seen on scales smaller than in $r_{21,n}$.
This is because the entire [\OIII] intensity is dominated by galaxies with $L_{\rm O_{III}} < 10^{42}~\rm erg~s^{-1}$ and these fainter galaxies are embedded in smaller ionized bubbles (see Fig. \ref{fig:map_21_gal}). 
We find that the shapes of the cross-power spectra are overall similar to $\Delta^2_{21,n}$ if the number density of galaxies with $L_{\rm O_{III}} > 10^{41}~\rm erg~s^{-1}$ is used. 

A large observation volume of intensity mapping may enable us to successfully detect a large-scale cross-correlation between 21-cm and intensity maps, but their noise levels can be estimated only if the detector, survey volume and depth are specified. 
In addition, a variety of foreground contamination needs to be considered.
Although accurate accounts of these factors are beyond the scope of the present paper, further studies are clearly warranted because metal-lines such as [\OIII] appear to provide a more promising probe than hydrogen/helium lines to identify early galaxy populations at $z>7$.

\section{DISCUSSION AND CONCLUSIONS}

\begin{figure}
\begin{center}
\includegraphics[width=7cm]{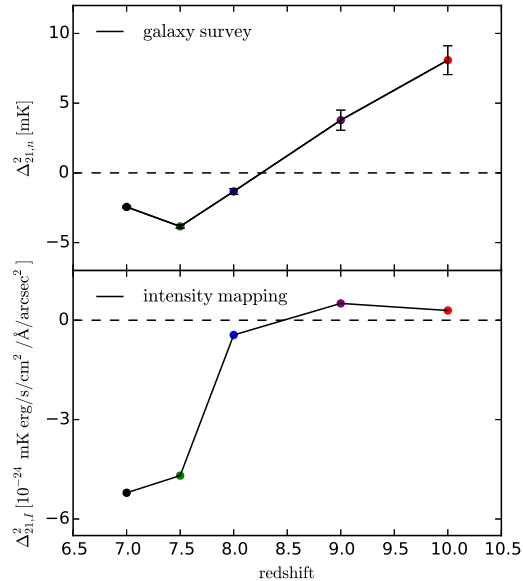}
\caption{
Top: the amplitude of the cross-power spectrum between 21-cm and the number density of galaxies with $L_{\rm O_{III}} > 10^{42}\rm ~ erg~s^{1}$ at $k = 0.1h~\rm cMpc^{-1}$ as a function of redshift. The errors are calculated assuming observations described in section 3 and with a bin width $\Delta \log k = 0.2$.
The red, purple, blue, green, and black points correspond to 
$(z, \overline{x}_{\rm HI}) = (10.0, 0.98)$, $(9.0, 0.94)$, $(8.0, 0.78)$, $(7.5, 0.46)$, and $(7.0, 0.23)$, respectively.
Bottom: the amplitude of the cross-power spectra between 21-cm and the intensity of [\OIII] emission as a function of redshift.
}
\label{fig:redshift_evolution}
\end{center}
\end{figure}

We have shown that the large-scale cross-power at $k \sim 0.1h~\rm cMpc^{-1}$ (and up to $k \sim 0.5h~\rm cMpc^{-1}$ depending on redshift) can be detected with SKA and a large volume galaxy survey or an intensity mapping.
A successful detection at various redshifts would be valuable to probe the ionization and thermal state of the IGM. 
In Fig. \ref{fig:redshift_evolution}  we show the redshift evolution of the cross-power amplitude at $k = 0.1h~\rm cMpc^{-1}$ using the number density of [\OIII] emitters (top) and [\OIII] intensity (bottom). 
For $\Delta^2_{21,n}$, we show the observational errors described in the previous sections.
From $z=10$ to 7.5, the IGM neutral fraction decreases rapidly from 1.0 to 0.5. 
As the temperature saturated regions and fully ionized regions grow in size, the sign of the cross-power spectrum changes from positive to negative.
This transition is a clear signature to identify the beginning of the EoR. 
Although the exact epoch of this transition depends on the population of ionizing and heating sources, 
we find that the transition occurs when $\overline{x}_{\rm HI} > 0.8$ unless heating gets saturated at a very late stage of the EoR.
To identify much earlier phases of the EoR, it is needed to detect and measure the cross-power spectra at smaller scales by reducing the observational errors. 

Overall, we argue that [\OIII] line emitters are excellent tracers 
of the underlying large-scale structure in the early universe
and thus offer a promising method to probe the process of early reionization.
In the future, the distribution of luminous [\OIII] 88 $\mu$m emitters at $z > 8$ can be probed, for instance, by a multi-object spectrograph DESHIMA/MOSAIC proposed to be installed on Large Submillimeter Telescope \citep{Kawabe16}.
The James Webb Space Telescope NIRCam can also be used to carry out a survey 
of optical [\OIII] 5007 \AA~line emitters at $z > 7$ with either its grism module or narrow band filters \citep{Moriwaki18}.
Finally, all-sky infrared intensity mapping to be carried out by SPHEREx may also allow us to study the large-scale clustering of [\OIII] emitters at $z > 8$. 
It is thus important to study further the distribution and the physical properties of
[\OIII] emitters and to understand their role in the process of cosmic reionization.

\section*{acknowledgments}

We thank 
Saleem Zaroubi, 
Robin Kooistra,
Keitaro Takahashi,
Kenji Kubota,
and Anastasia Fialkov
for fruitful discussions.
We also thank the anonymous referee for the helpful comments.
KM has been supported by the Grant-in-aid for the Japan Society for the Promotion of Science (19J21379) and by Advanced Leading Graduate Course for Photon Science (ALPS) of the University of Tokyo.

\bibliography{bibtex_library}

\label{lastpage}

\end{document}